\documentclass[aps,nofootinbib,superscriptaddress,12pt]{revtex4}
\usepackage{graphics}
\usepackage{graphicx}
\usepackage{amssymb,amsmath}

\begin{document}
\title{Search for the process $\mathbf{e^+e^- \to\eta}$}

\begin{abstract}
A search for the rare decay
$\eta \rightarrow e^+e^-$ is performed using the inverse process
$e^+e^- \rightarrow \eta$ in the decay mode
$\eta\to\pi^0\pi^0\pi^0 $. We analyze data with an integrated
luminosity of 654 nb$^{-1} $ accumulated at the VEPP-2000 $e^+e^-$ collider
with the SND detector at the center-of-mass energy
$E=m_\eta c ^2\approx 548 $ MeV, and set the upper limit
$ {\cal B} (\eta \rightarrow e^+ e^-) < 7 \times 10^{-7} $ at the 90\% 
confidence level.
\end{abstract}

\author{M.~N.~Achasov}
\author{A.~Yu.~Barnyakov}
\author{K.~I.~Beloborodov}
\author{A.~V.~Berdyugin}
\author{D.~E.~Berkaev}
\affiliation{Budker Institute of Nuclear Physics, SB RAS, Novosibirsk, 630090,
Russia}
\affiliation{Novosibirsk State University, Novosibirsk, 630090, Russia}
\author{A.~G.~Bogdanchikov}
\author{A.~A.~Botov}
\affiliation{Budker Institute of Nuclear Physics, SB RAS, Novosibirsk, 630090,
Russia}
\author{T.~V.~Dimova}
\author{V.~P.~Druzhinin}
\author{V.~B.~Golubev}
\author{L.~V.~Kardapoltsev}
\affiliation{Budker Institute of Nuclear Physics, SB RAS, Novosibirsk, 630090,
Russia}
\affiliation{Novosibirsk State University, Novosibirsk, 630090, Russia}
\author{A.~S.~Kasaev}
\affiliation{Budker Institute of Nuclear Physics, SB RAS, Novosibirsk, 630090,
Russia}
\author{A.~G.~Kharlamov}
\affiliation{Budker Institute of Nuclear Physics, SB RAS, Novosibirsk, 630090,
Russia}
\affiliation{Novosibirsk State University, Novosibirsk, 630090, Russia}
\author{I.~A.~Koop}
\affiliation{Budker Institute of Nuclear Physics, SB RAS, Novosibirsk, 630090,
Russia}
\affiliation{Novosibirsk State University, Novosibirsk, 630090, Russia}
\affiliation{Novosibirsk State Technical University, Novosibirsk, 630092, Russia}
\author{L.~A.~Korneev}
\affiliation{Budker Institute of Nuclear Physics, SB RAS, Novosibirsk, 630090,
Russia}
\author{A.~A.~Korol}
\affiliation{Budker Institute of Nuclear Physics, SB RAS, Novosibirsk, 630090,
Russia}
\affiliation{Novosibirsk State University, Novosibirsk, 630090, Russia}
\author{D.~P.~Kovrizhin}
\author{S.~V.~Koshuba}
\affiliation{Budker Institute of Nuclear Physics, SB RAS, Novosibirsk, 630090,
Russia}
\author{A.~S.~Kupich}
\affiliation{Budker Institute of Nuclear Physics, SB RAS, Novosibirsk, 630090,
Russia}
\affiliation{Novosibirsk State University, Novosibirsk, 630090, Russia}
\author{R.~A.~Litvinov}
\author{K.~A.~Martin}
\affiliation{Budker Institute of Nuclear Physics, SB RAS, Novosibirsk, 630090,
Russia}
\author{N.~A.~Melnikova}
\author{N. Yu. Muchnoi}
\affiliation{Budker Institute of Nuclear Physics, SB RAS, Novosibirsk, 630090,
Russia}
\affiliation{Novosibirsk State University, Novosibirsk, 630090, Russia}
\author{A.~E.~Obrazovsky}
\author{E.~V.~Pakhtusova}
\affiliation{Budker Institute of Nuclear Physics, SB RAS, Novosibirsk, 630090,
Russia}
\author{K.~V.~Pugachev}
\author{Yu.~A.~Rogovsky}
\author{A.~I.~Senchenko}
\author{S.~I.~Serednyakov}
\author{Z.~K.~Silagadze}
\affiliation{Budker Institute of Nuclear Physics, SB RAS, Novosibirsk, 630090,
Russia}
\affiliation{Novosibirsk State University, Novosibirsk, 630090, Russia}
\author{D.~N.~Shatilov}
\affiliation{Budker Institute of Nuclear Physics, SB RAS, Novosibirsk, 630090,
Russia}
\author{Yu.~M.~Shatunov}
\affiliation{Budker Institute of Nuclear Physics, SB RAS, Novosibirsk, 630090,
Russia}
\affiliation{Novosibirsk State University, Novosibirsk, 630090, Russia}
\author{D.~A.~Shtol}
\affiliation{Budker Institute of Nuclear Physics, SB RAS, Novosibirsk, 630090,
Russia}
\author{D.~B.~Shwartz}
\affiliation{Budker Institute of Nuclear Physics, SB RAS, Novosibirsk, 630090,
Russia}
\affiliation{Novosibirsk State University, Novosibirsk, 630090, Russia}
\author{I.~K.~Surin}
\author{ Yu.~V.~Usov}
\affiliation{Budker Institute of Nuclear Physics, SB RAS, Novosibirsk, 630090,
Russia}
\author{A.~V.~Vasiljev}
\author{V.~N.~Zhabin}
\author{V. V. Zhulanov}
\affiliation{Budker Institute of Nuclear Physics, SB RAS, Novosibirsk, 630090,
Russia}
\affiliation{Novosibirsk State University, Novosibirsk, 630090, Russia}

\maketitle

\section{Introduction}
This article is devoted to the search for the decay $\eta\to e^+e^-$ at the 
VEPP-2000 $e^+e^-$ collider. The experiment was proposed in Ref.~\cite{eta}.
For the measurement of the decay, the inverse reaction $ e^+e^-\to\eta $ 
is used.

\begin{figure}
\centering
\includegraphics[width=0.5 \textwidth]{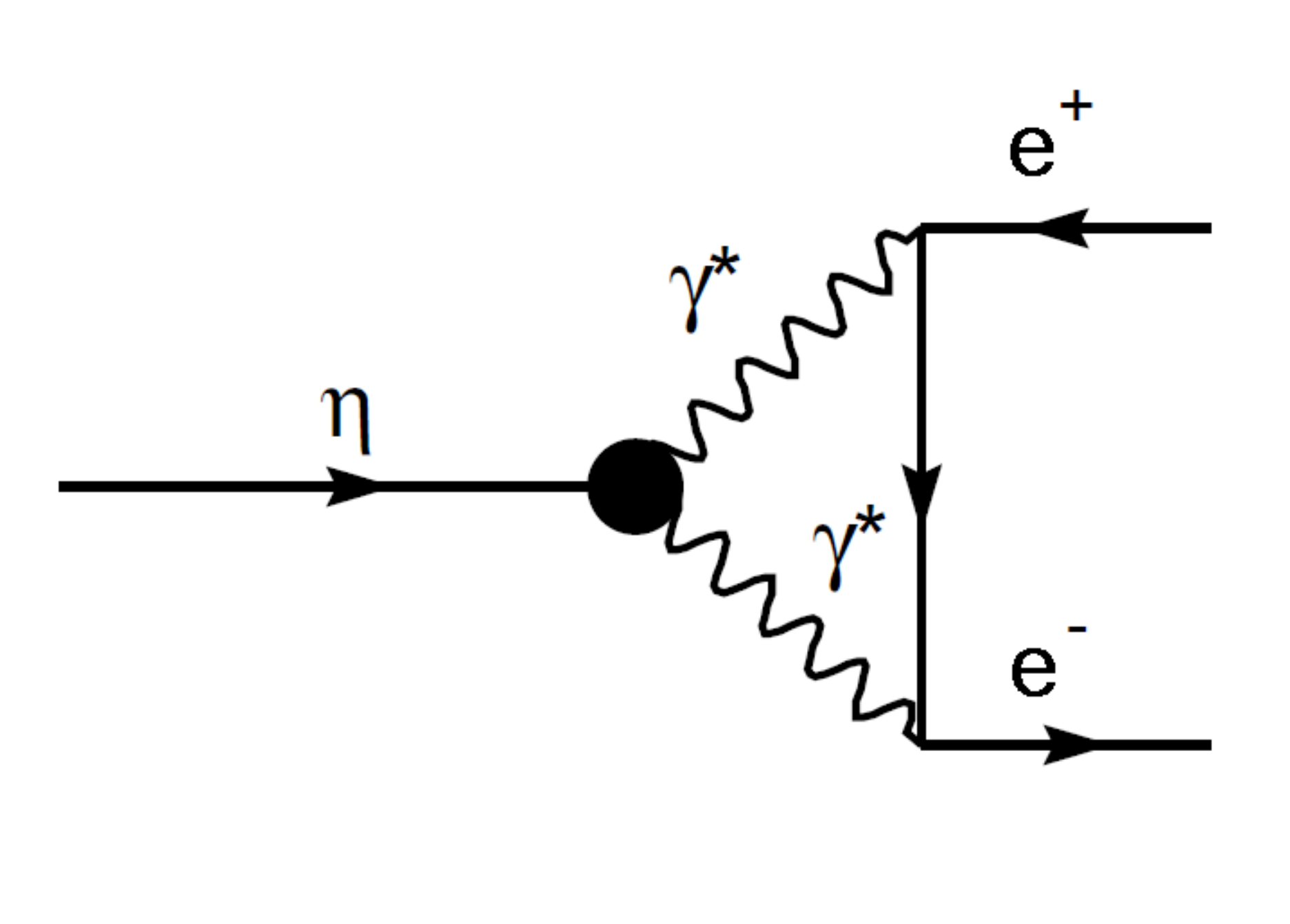}
\caption{The diagram for the $\eta\to e^+e^-$ decay.
\label{diag}}
\end{figure}

Decays of pseudoscalar mesons to lepton pairs $P\to l^+l^-$ ($l=e,\mu$) 
are rare. In the Standard Model, they proceed through the two-photon 
intermediate state, as shown in Fig.~\ref{diag}.
An additional suppression by a factor of $(m_l/m_P)^2$ arises from
the approximate helicity conservation. Thus, the width of the decay
$ P\to l^+l^- $ is less than the corresponding two-photon width
$ \Gamma (P \to \gamma\gamma )$ by a factor proportional to 
$ \alpha^2 (m_l/m_P)^2 $. Because of the small probability these decays are 
sensitive to contributions that are not described in the framework of the 
Standard Model~\cite{rare1,rare2}. It should be noted that the imaginary part
of the decay amplitude can be calculated from the width of the 
$P \to \gamma\gamma$ decay. This allows us to obtain a model-independent lower
boundary for the decay branching fraction, the so-called unitary 
limit~\cite{ul}. For the $\eta \rightarrow e^+ e^-$ decay it is equal to
${\cal B}^{\rm UL} (\eta \rightarrow e^+ e^-) = 1.78 \times 10 ^ {- 9} $.
It is expected that the total $\eta \rightarrow e^+ e^-$ branching fraction
exceeds the unitary limit by a factor of 2.5--3~\cite {th1,th2,th3}. 

The decay $\eta\to e^+e^-$ was not 
observed. The best upper limit on the decay branching fraction
$ {\cal B} (\eta \rightarrow e^+e^-)<2.3 \times 10^{-6} $ was set at
the HADES experiment~\cite{hades}. This paper presents the result of the
search for the decay $\eta \rightarrow e^+e^-$ performed in the SND 
experiment at the VEPP-2000 $e^+e^-$ collider.

\section{ Detector and Experiment }
The SND detector is described in detail in Refs.~\cite{det1,det2,det3,det4}. 
It is a non-magnetic detector, the main part of which is a three-layer 
spherical electromagnetic calorimeter consisting of 1640 NaI(Tl) crystals. 
The calorimeter covers a solid angle of 95\% of 4$\pi $. The energy and angular
resolutions for photons with energy $ E_\gamma $ are described by the 
following formulas:
\begin{eqnarray}
\sigma_{E_\gamma}/E_\gamma&=&4.2\%/\sqrt[4]{E_\gamma(\mbox{GeV})},\\
\sigma_{\theta,\phi} & = &0.82^{\circ}/\sqrt{E(\mbox{GeV})}.
\end {eqnarray}
Directions of charged particles are measured in a nine-layer drift chamber.
The calorimeter is surrounded by an iron absorber and a muon system.
In this analysis, the veto from the muon system is used for the suppression
of cosmic ray background.

The data used in this analysis were recorded
with the SND detector at the $e^+e^-$ collider VEPP-2000~\cite{vepp} 
in 2018 at the  center-of-mass (c.m.) energy $E$ near the $\eta$-meson mass 
$m_\eta c^2=547.862\pm0.017$ MeV~\cite{pdg}. The integrated luminosity
of 654 nb$^{-1}$ corresponding to this data set is measured using 
$e^+e^-\to \gamma\gamma$ events with an accuracy of
2\%~\cite{Achasov:2013btb}.

\section{Energy measurement } 
The $\eta$-meson width $\Gamma_\eta=1.31\pm0.05$ keV~\cite{pdg}
is much less than the c.m. energy spread $\sigma_E\approx 200$ keV.
In this case the visible cross section for the reaction $e^+e^-\to\eta$ is
proportional to the ratio $\Gamma_\eta/\sigma_E$. The knowledge of $\sigma_E$
is needed to measure the Born $e^+e^-\to\eta$ cross section and to extract the 
$\eta\to e^+e^-$ branching fraction. Also, it is necessary to be able to 
control the collider energy with an accuracy much better than $\sigma_E$.

At the VEPP-2000 collider there is a beam-energy-measurement system 
using the Compton back-scattering of laser photons on the electron
beam~\cite{emes}. The energy spectrum of scattered photons is measured 
by a high-purity germanium (HPGe) detector.
The energy $E_{\gamma,\rm{CBS}}$ corresponding to
the edge of the spectrum is related to the beam energy $E_b$:
\begin{equation}
E_{\gamma,\rm{CBS}}\approx \frac{4\omega_0E_b^2}{m_e^2c^4},
\end{equation}
where $\omega_0$ is the laser-photon energy, and $m_e$ is the electron mass.
The sharp edge of the Compton spectrum is smeared due to the beam-energy 
spread and the energy resolution of the HPGe detector.
The energy calibration of the detector and the measurement of its resolution
is performed using  well-known sources of $\gamma$-radiation~\cite{calib}.
In Ref.~\cite{emes} a system based on a CO laser with a wavelength
of 5.426463 $\mu$m is described. For this laser at $E_b=510$ MeV, the maximum 
energy of scattered photons is $E_{\gamma,\rm{CBS}}=0.90$ MeV, and the width
of the Compton spectrum edge due to the energy spread is 1.3 keV.
The latter value is comparable with the energy resolution of the HPGe
photon detector (0.9 keV). Below 500 MeV, the accuracy of the measurements
of the beam energy and especially the energy spread in the system with
the CO laser rapidly falls with decreasing the beam energy. Therefore, for 
experiments at $ E_b <500 $ MeV, a second, ytterbium fiber laser with a 
wavelength of 1.064966 $\mu$m, is used. The comparison of the beam-energy 
measurements with these two lasers has been performed at $E_b=512$ MeV. 
Two measurements are consistent within the statistical uncertainties (10 keV).

The systematic uncertainty of the beam-energy determination with the CO laser
was estimated in Ref.~\cite{emes} by comparison with the energy measurement
by the resonance depolarization method~\cite{rd} at $E_b=510$ and 460 MeV
($E_{\gamma,\rm{CBS}}=0.73$ MeV). It was estimated to be
$\Delta E_b/E_b=6\times 10^{-5}$. In the measurement with the ytterbium laser,
$E_{\gamma,\rm{CBS}}=0.73$ MeV corresponds to $E_b=200$ MeV. Therefore,
we conclude that the above estimation
$\Delta E_b/E_b=6\times 10^{-5}$ is valid for the ytterbium laser in
the beam-energy range from 200 to 500 MeV.
The systematic uncertainty in the c.m. energy determination at 
$E\approx m_\eta c^2$ is $33$ keV.

The energy spreads measured with the CO and ytterbium lasers
at $E_b=512$ MeV were also compared. They coincided within the 
10\% statistical errors. For the CO laser at $E_b=512$ MeV, the detector 
resolution and the beam-energy spread give comparable contributions to the 
width of the Compton-spectrum edge, whereas for the ytterbium laser
($E_{\gamma,\rm{CBS}}=4.6$ MeV) the detector resolution practically 
does not affect the beam-energy spread determination.
The coincidence of $\sigma_E$ obtained with the two lasers means that the 
detector resolution measured using radioactive sources is taken into account
correctly in the fit to the spectrum edge. At $E_b>500$ MeV the beam-energy 
spread can be evaluated using the formula~\cite{cmdetap}
\begin {equation}
\sigma_{E_b}=4.05\sigma_Z\sqrt{V_{\rm cav}E_b \sin(\arccos(63.2E_b^4V_{\rm cav}))},
\label {e1}
\end {equation}
where $\sigma_{E_b}$ is measured in keV, $E_b$ in GeV,  
the longitudinal beam size $\sigma_Z$ in mm, and the RF cavity
voltage $V_{\rm cav}$ in kV.
The length of the beam $\sigma_Z$ is measured using detected events of elastic
$e^+e^-$ scattering. Data collected at $E_b=511$, 550, 575, and 600 MeV
are used, in which the beam energy was measured with the ytterbium laser.
The beam-energy spread obtained by Eq.(\ref{e1}) is found to be
10-15\% lower than that measured on the Compton spectrum.
This difference (15\%) is taken as an estimation of the systematic
uncertainty on $\sigma_{E_b} $.

\begin{figure}
\centering
\includegraphics[width=0.7\textwidth]{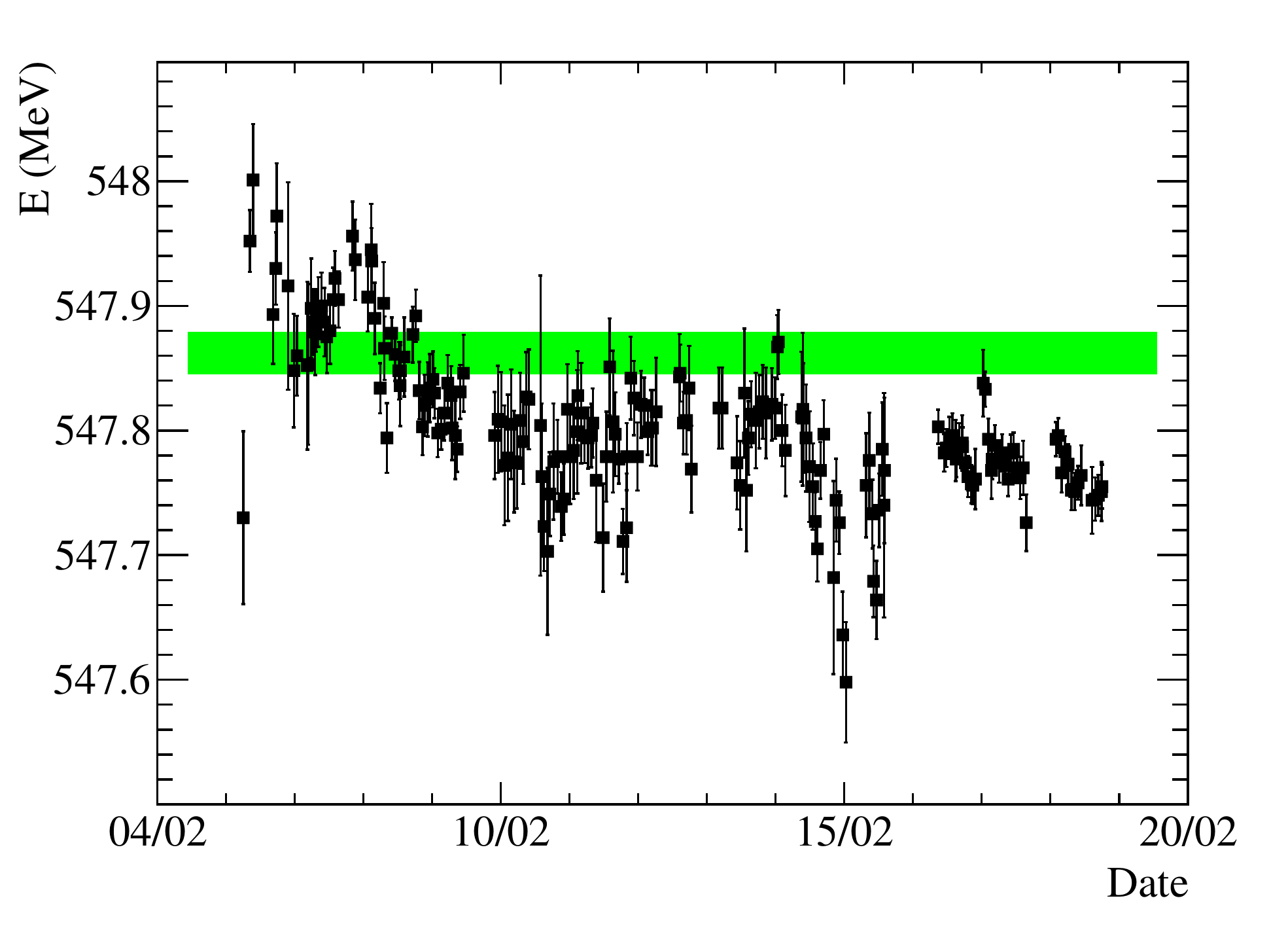}
\caption {The measurements of the c.m. energy during data taking.
The errors are statistical. The band indicates the $\pm 1\sigma$ range for
$E= m_{\eta} c^2 $~\cite{pdg}.
\label{fig2}}
\end{figure}

The beam-energy measurements is performed with a period of about 1 hour 
during data taking. These are presented in Fig.~\ref{fig2}.
The beam-energy spread was determined in every measurement. 
We do not observe nonstatistical deviations in the energy spread
during the experiment. Therefore, the average value $\sigma_E=226\pm7 \pm 34$
is used in the analysis, where the first error is statistical, and
the second is systematic.

\section {Calculation of the $e^+e^-\to \eta$ cross section}
The Born section for the reaction $e^+ e^-\rightarrow\eta$ is described by
the Breit-Wigner formula:
\begin{equation}
\sigma_{0} =\frac{4\pi}{E^2} {\cal B}(\eta \rightarrow e^+ e^-)
\frac{m^2_{\eta}\Gamma^2_{\eta }}{(m_{\eta}^2-E^2)^2+m_{\eta}^2\Gamma_\eta^2}.
\label{eq1}
\end{equation}

In analyses of experimental data it is necessary to take into account the radiative corrections,
arising, for example, from the emission of additional photons from the initial
state. To do this, we need to convolve the cross section (\ref{eq1}) with 
the so-called radiator function $W(s,x)$~\cite{rad1,rad2}
\begin {equation}
\sigma(s)= \int_{0}^{x_{max}}W(x,s)\sigma_{0}(s(1-x))dx, \label{eq2}
\end {equation}
where $s=E^2$, and $x_{max}=1-{(3m_{\pi^0})^2}/s$ for the decay 
$\eta\to3\pi^0$. The theoretical 
accuracy of the cross section~(\ref{eq2}) is better than 1\% ~\cite{rad1,rad2}.
For the unitary limit ${\cal B}^{\rm UL}(\eta \rightarrow e^+ e^-)=1.78\times 10^{-9}$, 
the Born cross section in the resonance maximum is $\sigma_0(m_\eta c^2)=29$ 
pb. The radiative corrections decrease this cross section up to 
$\sigma(m_\eta c^2)=14$ pb.

Since the $\eta$-meson width is much less than the c.m. energy spread, 
the cross section observed in the experiment is significantly smaller
than the cross section calculated above. It is calculated as a convolution 
of the cross section (\ref{eq2}) with a Gaussian function describing 
the energy spread:
\begin{equation}
\sigma_{\rm vis}(E_{0}) = \frac{1}{\sqrt{2\pi}\sigma_E}
\int\limits_{-\infty}^{+\infty}e^{-\frac{(E-E_{0})^2}{2\sigma^2_E}}\sigma(E) dE
\label{eq5}
\end{equation}
where $E_ {0}$ is the average collider c.m. energy.
For $\sigma_E=226 $ keV, $E_0=m_\eta c^2$, and
${\cal B}^{\rm UL}(\eta \rightarrow e^+ e^-)=1.78\times 10^{-9}$
the visible cross section (\ref{eq5}) is equal to
\begin{equation}
\sigma_{\rm vis}^{\rm UL}(m_\eta c^2)=72\pm 11\mbox{ fb}.
\label{xs-ul}
\end{equation}
The quoted uncertainty is due to the uncertainties in $ \sigma_E $ and 
$\Gamma_\eta $.
\begin{figure}
\centering
\includegraphics[width=0.7\textwidth]{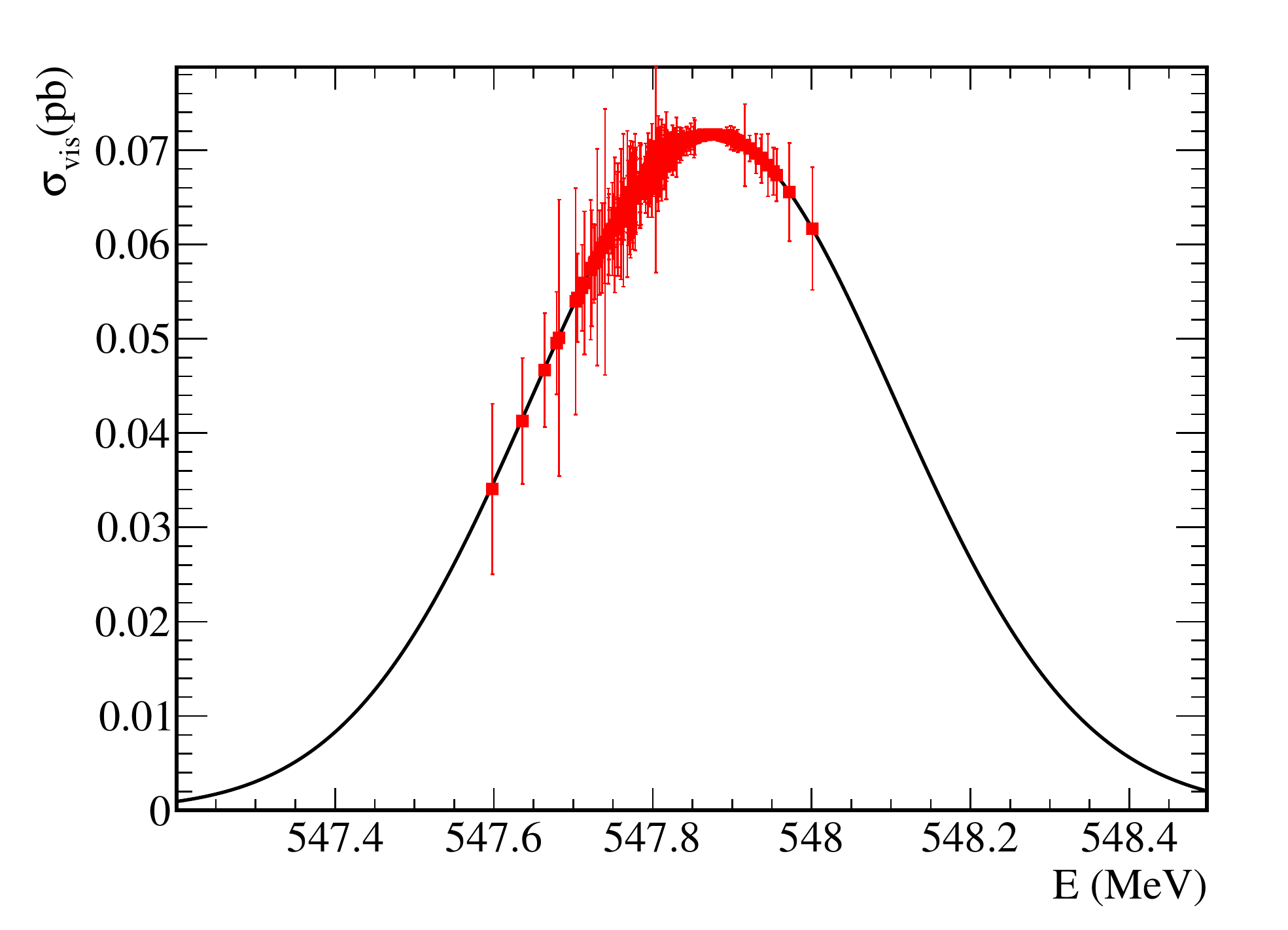}
\caption {The visible $e^+e^-\to \eta$ cross section evaluated for
${\cal B}^{\rm UL}(\eta \rightarrow e^+ e^-)=1.78\times 10^{-9}$ and
$\sigma_E=226 $ keV. The points with error bars represent the cross section 
values ​​at the energy points where data were recorded. The errors of the cross
section are determined by the statistical errors in the measurement of the beam
energy. \label {fig3}}
\end {figure}
 
The $\eta$-meson excitation curve obtained using Eq.(\ref{eq5}) is shown
in Fig.~\ref{fig3}. The points with error bars in Fig.~\ref{fig3} represent 
the cross section values ​​at the energy points, where data were recorded. 
The errors of the cross section are determined by the statistical errors in 
the measurement of the beam energy.
The expected $\eta$-meson production cross section is calculated as follows:
\begin{equation}
\sigma_{\rm vis}^{\rm UL} = \frac{\sum_i L_i\sigma(E_i)}{\sum_i L_i}=65\pm 9\mbox{ fb},
\label{eq50}
\end{equation}
where $ L_i $ is the integrated luminosity for the $i$-th energy point $ E_i $,
$\sigma(E_i)$ is the cross-section calculated using Eq.(\ref{eq5}).
The error of $\sigma_{\rm vis}^{\rm UL}$ includes contributions from 
the statistical errors in the beam-energy measurements (0.4 fb), uncertainties 
on $\sigma_E$ (7.9 fb), $\Gamma_\eta$ (0.3 fb), and $m_\eta $ (0.2 fb),
and the systematic uncertainty of the energy measurement (3 fb).

\section{\bf\boldmath Event selection}
The preferred $\eta$ decay mode for the search for the process 
$e^+e^-\to\eta$ with SND, for which physical background is small~\cite{eta}, 
is $\eta \rightarrow \pi^0\pi^0\pi^0\to 6\gamma$.
The main source of the background is cosmic rays. We select events
with exactly six photons and energy deposition in the calorimeter greater
than $0.6E$. Background from events with charged particles
is rejected by the requirement that the number of fired wires in the drift 
chamber is less than four. Cosmic-ray background is suppressed by the veto 
from the muon detector.

\begin {figure}
\centering
\includegraphics [width=0.7 \textwidth] {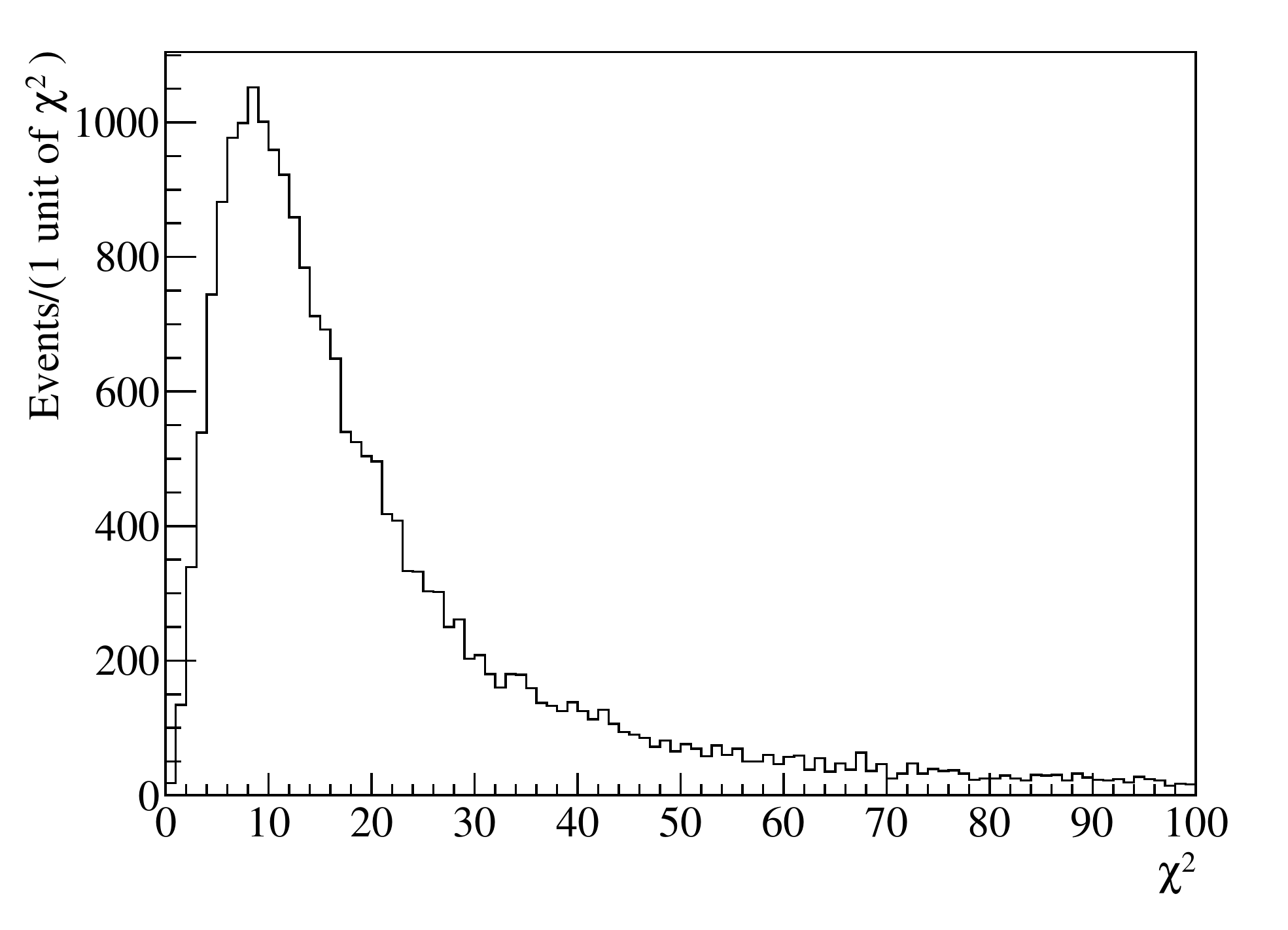}
\caption {{\rm Figure 2}. The $\chi^2$ distribution for
simulated $e^+e^-\rightarrow\eta$ events 
passed the selection criteria of six-photon events.
\label {fig20}}
\end {figure}

For the events passing the preliminary selection, a kinematic fit to
the hypothesis $e^+e^-\rightarrow\pi^0\pi^0\pi^0\rightarrow 6\gamma $ is 
performed with a requirement of total energy and momentum conservation and 
a condition that the invariant masses of the 3 pairs of photons are equal to 
the $\pi^0$ mass. The invariant mass of the $\pi^0$ candidate is required to
be in the range $m_{\pi^0}\pm 50$ MeV/$c^2$.
The quality of the kinematic fit is characterized by the $\chi^2$ parameter.
During the fit, all possible combinations of two-photon pairs are checked 
and a combination with the smallest value of $\chi^2$ is selected.
The $\chi^2$ distribution for simulated $e^+e^-\rightarrow\eta$ events
passed the selection conditions described above
is shown in Fig.~\ref{fig20}. The condition $\chi^2<100$ is used.
 
The detection efficiency for $ e^+e^-\rightarrow \eta $
events determined using simulation is equal to 
$\varepsilon = (14.1 \pm 0.7)\%$. 
The quoted error is systematic.
It is estimated using results of Ref.~\cite{Achasov:2013btb},
where data and simulated $\chi^2$ distributions 
were compared for five-photon events from the process
$e^+e^- \to \omega\pi^0 \to \pi^0\pi^0\gamma$.
 
No signal events passed the selection criteria described above are found 
in the data sample recorded with the SND detector at $E\approx m_{\eta}c^2$. 
 
The visible cross section for the process $ e^+e^- \rightarrow \eta $ and
the $\eta \rightarrow e^+e^-$ branching fraction are determined as
\begin {equation}
\sigma_{\rm vis} = \frac{N_s}{\varepsilon L},\;\;
{\cal B}(\eta \rightarrow e^+ e^-)={\cal B}^{\rm UL}(\eta \rightarrow e^+ e^-)
\frac{\sigma_{\rm vis}}{\sigma_{\rm vis}^{\rm UL}},
\label {eq8}
\end {equation}
where $ N_s $ is the number of selected events, $ \varepsilon $ is
the detection efficiency, and $ L $ is the integrated luminosity. 
Since no events of the process under study are found, we set the upper limit
on the branching fraction
\begin {equation}
{\cal B}(\eta \rightarrow e^+ e^-) < 7 \times 10^{-7}
\label {eq9}
\end {equation}
at the 90\% confidence level~\cite{barlow}. This result is more than 3 times
lower than the previous limit 
${\cal B}(\eta \rightarrow e^+ e^-)<2.3 \times 10^{-6}$~\cite{hades}.

\section{\bf\boldmath Summary}
The search for the process $ e^+ e^- \rightarrow \eta$ has been
carried out with the SND detector at the VEPP-2000 $e^+e^-$ collider 
in the decay mode $\eta\rightarrow \pi^0\pi^0\pi^0$. No candidate
events for the process $e ^+ e^-\rightarrow \eta$ has been found.
Since the visible $e ^+ e^-\rightarrow \eta$ cross section is proportional
to the branching fraction ${\cal B}(\eta \rightarrow e^+ e^-)$, 
the upper limit has been set
\begin{equation}
{\cal B}(  \eta \rightarrow e^+ e^- )<7 \times 10^{-7}
\label{eq90}
\end{equation}
at the 90\% confidence level.

\section{\bf\boldmath Acknowledgments}
Part of this work related to the beam energy measurement and the photon reconstruction 
algorithm in the electromagnetic calorimeter for multiphoton events is supported by the 
Russian Science Foundation (project No. 14-50-00080).

\end{document}